\documentclass[a4paper,11pt]{article}
\usepackage{pos}

\title{Opportunities at the Sanford Underground Research Facility}

\author*[a]{Jaret Heise}

\affiliation[a]{Sanford Underground Research Facility,\\
  630 East Summit Street, Lead, USA}

\emailAdd{jaret@sanfordlab.org}

\abstract
{The Sanford Underground Research Facility (SURF) has been operating for more than 15 years as an international facility dedicated to advancing compelling multidisciplinary underground scientific research in rare-process physics, as well as offering research opportunities in other disciplines. SURF laboratory facilities include a Surface Campus as well as campuses at the 4850-foot level (1490~m, 4300~m.w.e.\@) that host a range of significant physics experiments, including the LUX-ZEPLIN (LZ) dark matter experiment and the {\sc Majorana Demonstrator} neutrinoless double-beta decay experiment. The CASPAR nuclear astrophysics accelerator completed the first phase of operation and is planning for the second phase beginning in 2024. SURF is also home to the Long-Baseline Neutrino Facility (LBNF) that will host the international Deep Underground Neutrino Experiment (DUNE). SURF offers world-class service, including an ultra-low background environment, low-background assay capabilities, and electroformed copper is produced at the facility. SURF is preparing to increase underground laboratory space. Plans are advancing for construction of new large caverns (nominally 100m L x 20m W x 24m H) on the 4850L (1485~m, 4100~m.w.e.\@) on the timeframe of next-generation experiments ($\sim$2030). SURF plans to leverage existing advisory and community committees as well as engage the underground science community to inform plans for future laboratory space.}

\FullConference{%
  XVIII International Conference on Topics in Astroparticle and Underground Physics \\
  28 August - 1 September 2023 \\
  University of Vienna, Vienna, Austria
}

\renewcommand{\logo}{\relax}

\begin{document}
\maketitle

\section{Introduction}
The Sanford Underground Research Facility (SURF) is an international facility dedicated to advancing world-class multidisciplinary science and inspiring learning across generations~\cite{Heise:2022iaf}. The unique underground environment at SURF attracts experiments with the potential to revolutionize our understanding of the universe, including the nature of dark matter, the properties of neutrinos and topics related to nuclear astrophysics such as the synthesis of atomic elements within stars.

With strong support from the scientific community as well as federal, state and private (T. Denny Sanford) funding, SURF has been operating as a dedicated research facility for more than 16 years. Since Fall 2019, SURF operation has been funded by the U.S.\ Department of Energy’s (DOE) Office of Science through a Cooperative Agreement; at a future date SURF anticipates being designated as a national DOE User Facility. The SURF organization comprises approximately 200 full/part-time staff in 11 departments and 6 offices.

\section{Facilities and Support}
SURF property comprises approximately 1~km$^{2}$ on the surface and more than 31~km$^{2}$ underground. Of a total of 600~km of tunnels that extend 2450~m below ground, access to roughly 35~km is currently maintained, including seven key elevations identified for science activities.

Two underground research campuses are located on the 4850-foot level of the facility. The Davis Campus (near the Yates Shaft) has a total footprint of 3017~m$^{2}$ and includes a stainless steel tank that can be used for shielding (7.6~m diameter, 6.4~m high). The Ross Campus (near the Ross Shaft) consists of four areas with a total footprint of 2653~m$^{2}$, with two spaces currently configured as laboratories. Cleanrooms and radon-reduction systems support activities both on the surface and underground. Due to Long-Baseline Neutrino Facility (LBNF) construction, Ross Campus laboratories were temporarily mothballed in 2021, and activities are expected to resume in 2024.

The Black Hills State University (BHSU) Underground Campus (BHUC) provides an international capability for low-background assays, supporting current and future SURF experiments as well as the broader underground science community~\cite{Tiedt:2023gld}. Production assays currently provide sensitivities as low as 100~$\mu$Bq/kg U/Th; dual-crystal systems may achieve 10~$\mu$Bq/kg U/Th.

SURF is one of only a few laboratories in the world where underground copper electroforming is currently performed (average uranium and thorium decay-chain backgrounds $\leq$~0.1~$\mu$Bq/kg).

\section{Science Program}
Building on the legacy of the Ray Davis chlorine solar-neutrino experiment, 65 groups have conducted underground research at SURF since 2007. A total of 32 research programs are ongoing, with 400 active researchers from a pool of over 2235 collaborators (including the Deep Underground Neutrino Experiment (DUNE)), representing efforts from 294 institutions in 39 countries. So far in 2023, there have been expressions of interest from 25 research groups, which sustains a trend of growing interest over the past several years.

SURF hosts the LUX-ZEPLIN (LZ) experiment that is searching for dark matter using 10~tonnes of xenon~\cite{LZ-TAUP2023-Composite}, as well as the {\sc Majorana Demonstrator} experiment that completed its neutrinoless double-beta decay search using germanium in 2021~\cite{MJD-TAUP2023-Composite-1} and is now performing a rare decay search with Ta-180m~\cite{MJD-TAUP2023-Composite-2}. The Compact Accelerator System for Performing Astrophysical Research (CASPAR) nuclear astrophysics experiment~\cite{Dombos:2022bph} completed the first phase of operation in 2021 and expects to resume operations in 2024. The DUNE liquid-argon experiment at the 4850L LBNF Campus~\cite{DUNE-TAUP2023-Composite} will begin collecting physics data in the 2028/2029 timeframe, ultimately investigating neutrino properties (oscillations, CP violation, mass hierarchy), nucleon decay and supernovae. 

\section{Future Plans}
Construction for LBNF/DUNE is underway at SURF. The excavation phase for two large caverns (each 145~m L $\times$~20~m W~$\times$ 28~m H) and a utility cavern is expected to be completed by mid-2024. Once completed, LBNF/DUNE will comprise a total of 22,922~m$^{2}$ (242,190~m$^{3}$).

\begin{figure}[!htbp]
  \centering
  \includegraphics*[height=0.5\columnwidth,angle=0]{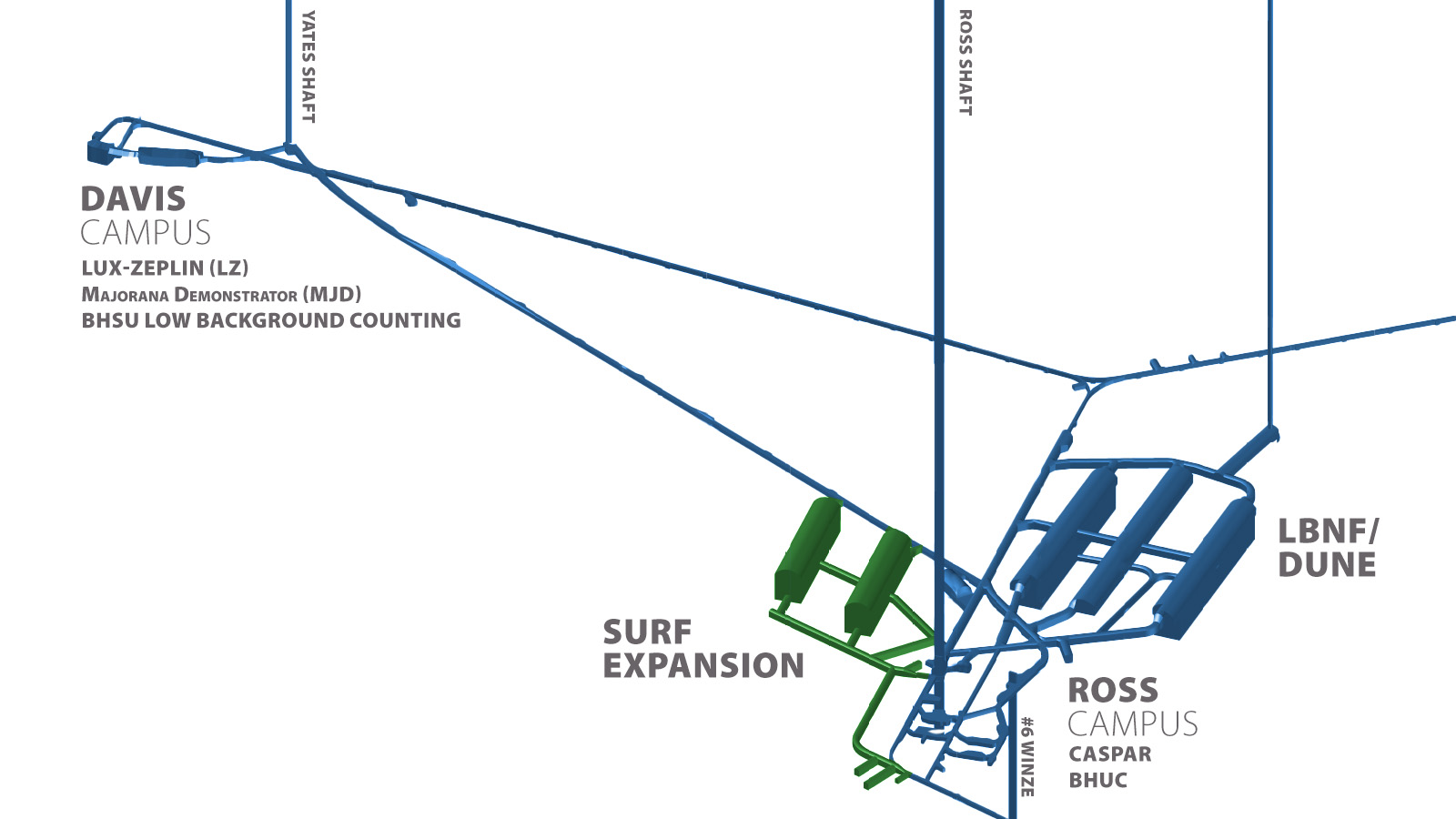}
  \caption{\label{fig:expansion} SURF current and proposed underground laboratory space. Two new caverns in green are shown (100~m L $\times$~20~m W~$\times$ 24~m H) with a rock overburden similar to existing laboratories (1485~m, 4100~m.w.e.\@).}
\end{figure}

As part of SURF's strategic plan, preparations are advancing to increase underground laboratory space on the 4850L. A 2021 feasibility study confirmed that cavern excavation near the Ross Shaft could be performed without interference to LBNF/DUNE. With a generous appropriation of \$13M from the South Dakota legislature in March 2023, design is underway to support initial excavation starting in early 2024. Up to two new caverns are envisioned using private funding, each with a cross-section of 20~m wide $\times$ 24~m high and up to 100~m in length as shown in \autoref{fig:expansion}. Excavation for two 100-m caverns is estimated to take 2.5~years, aligning with the timeframe for construction of next-generation experiments ($\sim$2030). Based on the current state of the design, the expansion will consist of 7,175~m$^{2}$ (110,649~m$^{3}$) total space and 4,047~m$^{2}$ (94,607~m$^{3}$) science space.

\section{Summary}
SURF is a deep underground research facility dedicated to scientific uses that has been operating for 16 years, offering world-class service and a proven track record of enabling experiments to deliver high-impact science across diverse scientific communities. Research activities, including dark matter, neutrinoless double-beta decay and nuclear astrophysics, are supported at surface and underground facilities. Electroformed copper production and assay capabilities are also available. 

In addition to the existing science program as well as hosting LBNF/DUNE, SURF is eager to host future experiments, and to that end, expansion at SURF is on the horizon. In addition to the existing two 4850L campuses and the upcoming LBNF Campus for DUNE, SURF is actively preparing to increase underground laboratory space on the 4850L. A mixture of federal, state and private funding will allow phased development of underground space aligned with needs for next-generation neutrino and dark matter projects.

\section*{Acknowledgments}
This material is based upon work supported by the U.S.\ Department of Energy Office of Science under award number DE-SC0020216.

\bibliographystyle{utphys}
\bibliography{main}

\end{document}